\documentclass[conference,twocolumn,a4paper]{IEEEtran}
\usepackage{amssymb,amsmath,amsthm} 
\usepackage{cite} 
\usepackage{psfrag}
\usepackage{here}
\usepackage{algorithm}
\usepackage{algpseudocode}
\algblock{Output}{EndOutput}
\algnotext{EndOutput}
\algnewcommand{\Initialize}[1]{%
	\State \textbf{Initialize:}
	\Statex \hspace*{\algorithmicindent}\parbox[t]{.8\linewidth}{\raggedright #1}
}
\usepackage[left=1.57cm, right=1.58cm, top=1.71cm,bottom=4.31cm]{geometry}

\usepackage{graphicx,psfrag}
\usepackage{fancyhdr,afterpage,lipsum}
\usepackage{booktabs}
\usepackage{tabularx}
\usepackage{multicol}
\usepackage{multirow}
\usepackage{color}
\usepackage{subcaption}
\usepackage{url} 
\usepackage[latin1]{inputenc} 

\theoremstyle{remark}  

\title{Multi-Static ISAC in Cell-Free Massive MIMO: Precoder Design and Privacy Assessment}

\author{\IEEEauthorblockN{Isabella W. G. da Silva, Diana P. M. Osorio, and Markku Juntti}
\IEEEauthorblockA{\textit{Centre for Wireless Communications, University of Oulu, P.O.Box 4500, FI-90014, Finland} \\
Emails: \{isabella.dasilva, diana.moyaosorio, markku.juntti\}@oulu.fi}}

\begin{document}
\maketitle
\vspace{-1cm}
\begin{abstract}
A multi-static sensing-centric integrated sensing and communication (ISAC) network can take advantage of the cell-free massive multiple-input multiple-output infrastructure to achieve remarkable diversity gains and reduced power consumption. While the conciliation of sensing and communication requirements is still a challenge, the privacy of the sensing information is a growing concern that should be seriously taken on the design of these systems to prevent other attacks. This paper tackles this issue by assessing the probability of an internal adversary to infer the target location information from the received signal by considering the design of transmit precoders that jointly optimizes the sensing and communication requirements in a multi-static-based cell-free ISAC network. Our results show that the multi-static setting facilitates a more precise estimation of the location of the target than the mono-static implementation. 
\end{abstract}
\section{Introduction}
The sixth generation (6G) of wireless communications is expected to heavily influence every aspect of the Society in 2030. Several applications envisioned for 6G will be supported by sensing capabilities, with the network acting as a sensor. For this to become reality, the new paradigm of integrated sensing and communications (ISAC) plays an important role by allowing an intelligent share of wireless resources and the support for a number of use cases such as autonomous vehicles and smart homes~\cite{9737357}.  

So far, the research on ISAC systems has mainly focused on mono-static sensing. However, the employment of mono-static sensing requires the co-located sensing transmitter and receiver to be full-duplex. Hence, multi-static sensing ISAC systems, in which there are multiple non-colocated transmitters and receivers for sensing, can provide diversity gain while avoiding the need for full-duplex nodes~\cite{Art:Emil}. For instance, in~\cite{Art:Emil}, Behdad \textit{et al.} assessed the probability of detection of a single target in a cell-free massive multiple-input multiple-output (MIMO) ISAC system where several ISAC transmitters and sensing receivers are deployed to detect the target and to transmit data to multiple communication users (UEs). Moreover, in~\cite{9842350}, Huang \textit{et al.} proposed a coordinated power control design for a networked ISAC system aiming to maximize the signal-to-interference-plus-noise ratio (SINR) at the UEs and maximize the Cr\'{a}mer-Rao lower bound (CRLB) of the target location estimate. Both works demonstrated that in comparison to communication-centric designs, a cell-free massive MIMO ISAC design requires less transmit power to attain more accurate detection.

Nevertheless, the design of joint precoders for sensing and communications employed in most of the works with mono-static or multi-static ISAC systems considered the simultaneous transmission of individual radar and communication waveforms~\cite{9124713,demirhan2023cell}. Thus, the transmit precoder contains information about the target location and the data intended for the UEs. Indeed, the advancements in ISAC design have exacerbated security and privacy concerns since the communication information could be exposed to untrusted targets, and malicious UEs could try to infer the position of targets. In this sense, security and privacy aspects have recently started to emerge as a crucial part of the design of ISAC. For instance, the works in \cite{art:su2021, 10104579} and \cite{art:ren} have focused on the design of beamformers with secrecy requirements to prevent untrusted targets from eavesdropping information from UEs. In~\cite{art:su2021}, Su \textit{et al.} proposed a beamforming design to minimize the signal-to-noise ratio (SNR) at a potential malicious target constrained to minimal SINR requirements for the UEs. In~\cite{10104579}, the transmit beamforming is obtained via a weighted optimization between radar estimation CRLB and the communication secrecy rate. Also, in~\cite{art:ren}, Ren \textit{et al.} proposed a beamforming design to minimize the beampattern matching error constrained to secrecy rates requirements in a network with multiple targets, with some of them assumed as malicious.

On the other hand, the privacy of the sensing information remains barely explored while being an important source of concern as it can trigger further attacks. In our earlier work~\cite{dasilva2023privacy}, we addressed this aspect by evaluating the capability that an internal adversary of the network will have to infer the angular position of a target by a UE that acts as an adversary. Results demonstrated that if the adversary is capable of inferring the transmit precoder, the detection surpasses fifty percent of the cases, which entails a serious privacy breach for ISAC systems. 

Accordingly, recognizing the benefits of multi-static sensing and the seriousness of possible privacy leakages in ISAC systems, in this work we investigate a multi-static sensing-centric ISAC network in a cell-free massive MIMO scenario, with multiple MIMO UEs and a point-like target. In this scenario, one of the UEs acts maliciously by trying to estimate the position of the target from the observations of the received signal. The main contributions of this work are as follows: i) we propose a transmit precoder design to maximize the sensing SINR constrained to a minimal SINR for the UEs and a transmit power limit requirement; ii) a replica of the transmit beampattern from each transmitter access point (AP) is computed based on an expectation-maximization (EM) approach. 

The remainder of this paper is organized as follows: Section II describes the system and channel models for the considered ISAC system. In Section III, the transmit precoder is designed. Section IV describes the adversary model. Section V illustrates the numerical results and discussion, and Section VI concludes the paper.
\section{System Model}\label{sec:model}
\vspace{-0.3cm}

\begin{figure}[h]
    \centering\includegraphics[scale=0.2]{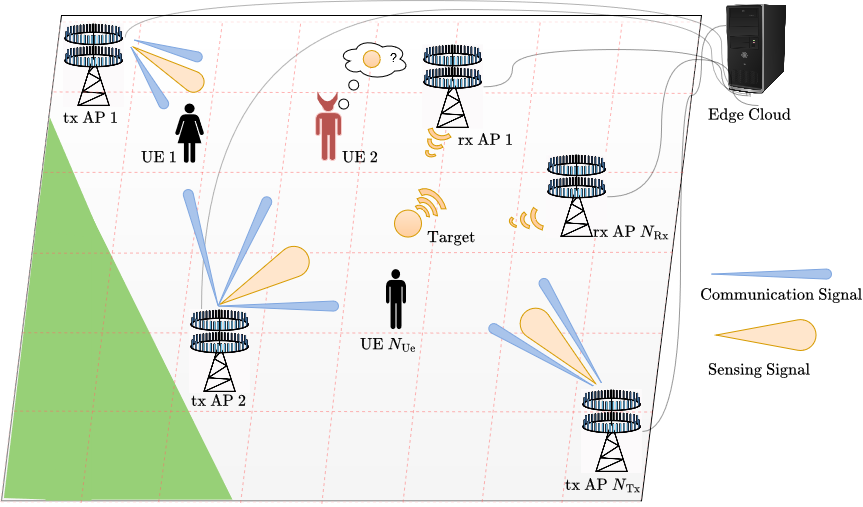}
    \caption{System Model}
    \label{fig:systemmodel}
  \vspace{-0.3cm}
\end{figure}

Consider the ISAC system as illustrated in Fig.~\ref{fig:systemmodel}, where $N_{\mathrm{Tx}}$ APs are jointly serving $N_{\mathrm{Ue}}$ UEs equipped with $M_{\mathrm{Ue}}$ antennas, while also detecting a point-like target. In this system, one of the UEs acts as an adversary and intends to discover the location of the target. The other $N_{\mathrm{Rx}}$ APs are acting as sensing receivers that simultaneously sense the location of the target. The transmit and receive APs are equipped with $M_{\mathrm{AP}}$ half-wavelength-spaced isotropic antennas deployed as an uniform linear array (ULA). It is further assumed that all APs are connected to the edge cloud, where the processing is done in a centralized manner, and are fully synchronized. Similar to~\cite{Art:Emil}, we assume that the transmitted signal by the $k$th transmit AP is given by a weighted sum of communication symbols and sensing signals. Hence, at time instance $n$, the transmitted signal $\mathbf{x}_k[n]$$\in$$\mathbb{C}^{M_{\mathrm{AP}}}$ is written as  
\vspace{-0.2cm}
\begin{align}
    \mathbf{x}_k[n]\!=\!\sum_{i=1}^{N_{\mathrm{Ue}}} \mathbf{w}_{i,k} s_i[n]\!+\!\mathbf{w}_{t,k} s_t[n]\!=\!\mathbf{W}_k \mathbf{s}[n],\nonumber\\ n = 0,..., N\!-\!1, k = 1,..., N_{\mathrm{Tx}},
\end{align}
where $\mathbf{s}[n]$$=$$[s_1[n],...,s_{N_{\mathrm{Ue}}}[n],s_{t}[n]]^T$ is the $(N_{\mathrm{Ue}}\!+\!1)$$\times$$1$ vector containing the $N_{\mathrm{Ue}}$$\times$$1$ parallel communications symbols intended to the $N_{\mathrm{Ue}}$ users plus the sensing signal, which is independent of the UE's data signals. Also, $\mathbf{w}_{i,k}$ and $\mathbf{w}_{t,k}$ are the $M_{\mathrm{AP}}\!\times\! 1$  transmit precoder vector of the $k$th transmitter AP for the $i$th user and for the sensing of the target, respectively. In addition, the users are assumed to employ a receive beamformer $\mathbf{u}_{i}$, with size $M_{\mathrm{Ue}}$$\times$$1$, to estimate the transmitted data stream. Accordingly, the estimated data stream at the $i$th user at time $n$ is given by
\begin{align}\label{eq:signali}
    y_{i}[n]&= \mathbf{u}_i^H\left(\sum_{k=1}^{N_{\mathrm{Tx}}}\mathbf{H}_{i,k}\mathbf{W}_{k}\mathbf{s}[n] +\mathbf{n}_i[n]\right),
\end{align}
where $\mathbf{H}_{i,k}$ is the $M_{\mathrm{Ue}}$$\times$$M_{\mathrm{AP}}$ channel coefficient matrix between the $k$th AP and UE $i$. It is assumed that there is no dominant line-of-sight (LOS) component between APs and UEs, so $\mathbf{H}_{i,k}$,   undergoes Rayleigh block fading, for all $i \in 1, ..., N_{\mathrm{Ue}}$. Moreover, $\mathbf{n}_i$ is the noise component at the $i$th user, modeled as signal-independent, zero-mean, additive white Gaussian noise (AWGN) with variance $\sigma^2_i\mathbf{I}$. Following, the SINR at UE $i$ given the transmission of AP $k$ is written as
\begin{align}
    \gamma_{i,k}\!=\!\frac{|\mathbf{u}_i^H\mathbf{H}_{i,k}\mathbf{w}_{i,k}|^2}{\sum_{l=1 \atop l \neq i}^{N_{\mathrm{Ue}}}\!|\mathbf{u}_i^H\mathbf{H}_{i,k}\mathbf{w}_{l,k}|^2\!+\!|\mathbf{u}_i^H\mathbf{H}_{i,k}\mathbf{w}_{t,k}|^2\!+\!\sigma^2_i||\mathbf{u}_i||^2}.
\end{align}
On the other hand, the signal received by the $r$th receiver AP at the time instant $n$ is written as
\begin{align}
\mathbf{y}_r[n]\!=\!\!\!\sum_{k=1}^{N_{\mathrm{Tx}}}\!\!\alpha_{r,k} \sqrt{\beta_{r,k}}\mathbf{a}(\phi_r)\mathbf{a}^T(\phi_k)\mathbf{x}_k[n] \!+\!\mathbf{n}_r[n],
\end{align}
where $\alpha_{r,k}$ is the bi-static unknown radar cross section (RCS) of the target through the reflection path from transmitter AP $k$ to the receiver AP $r$. Herein, as in~\cite{Art:Emil}, it is assumed that the RCS follows the Swerling-I model, so that $\alpha_{r,k}$ is constant throughout a collection of consecutive sensing signals and follows the distribution $\alpha_{r,k}$$\sim$$\mathcal{CN}(0,\sigma_{r,k}^2)$. Also, $\phi_i$ is the azimuth angle from the target position to the $i$th AP, with $i \in \{k = 1, \dots, N_{\mathrm{Tx}},r = 1, \dots, N_{\mathrm{Rx}}\}$. 
Accordingly, $\mathbf{a}(\phi) \in \mathbb{C}^{M_{\mathrm{AP}}}$ is the antenna array steering vector, given by
\begin{align}
    \mathbf{a}(\phi)\!\!=\!\!\left[
        1, e^{j\pi\sin(\phi)}, \dots, e^{j(M_{\mathrm{AP}}\!-\!1)\pi\sin(\phi)}\right]^T\!\!\!\!.
\end{align}
Finally, $\mathbf{n}_r[n]$$\sim$$\mathcal{CN}(\mathbf{0},\sigma_n^2\mathbf{I}_{M_{\mathrm{AP}}})$ is the noise component at the $r$th receiver AP, and $\beta_{r,k}$ is the channel gain of the path between the $k$th transmitter AP to the target and from the target to the $r$th receiver AP, which can be calculated as~\cite{Art:Emil} 
\begin{align}
    \beta_{r,k} = \frac{\lambda_c^2}{(4\pi)^3d_{t,k}^2d_{t,r}^2},
\end{align}
where $\lambda_c$ is the carrier wavelength, and $d_{t,i}$ is the distance between the target location and the AP i, with $i \in \{k = 1, \dots, N_{\mathrm{Tx}},r = 1, \dots, N_{\mathrm{Rx}}\}$. Moreover, the possible clutter caused by permanent or temporary objects is neglected. Accordingly, the sensing SINR, $\gamma_t$, is given by
\begin{align}\label{eq:gammat}
    \gamma_t\!=&\zeta\!\Bigg(\!\sum_{n=1}^{N-1}\!\!\mathbf{s}^H\![n]\!\Bigg(\!\sum_{r=1}^{N_{\mathrm{Rx}}}\! \sum_{k=1}^{N_{\mathrm{Tx}}}\! \sum_{j=1}^{N_{\mathrm{Tx}}}\! \varphi_{r,k,j}\mathbf{W}_k^H\mathbf{A}_{k,j}\mathbf{W}_j\!\Bigg) \!\mathbf{s}[n]\!\Bigg)\!,
\end{align}
with $\varphi_{r,k,j} = \sqrt{\beta_{r,k} \beta_{r,j}}\mathbf{a}^H\left(\phi_r\right)\operatorname{cov}\left(\alpha_{r, j}, \alpha_{r, k}\right) \mathbf{a}\left(\phi_r\right)$, $\mathbf{A}_{k,j}$$=$$\mathbf{a}^*\!\left(\phi_k\right)\mathbf{a}^T\!\left(\phi_j\right)$, and $\zeta$$=$$1/(N\!-1\!)M_{\mathrm{AP}}N_{\mathrm{Rx}}\sigma_n^2$.
\section{Precoder Design}
For the design of the ISAC transmit precoding, the goal is to maximize the sensing SINR in \eqref{eq:gammat} under transmit power and communication quality of service (QoS) constraints, which can be formulated as
\begin{subequations}
\begin{align}
\mathcal{P}: \max_{\mathbf{W},\mathbf{u}} &  \hspace{3mm}\gamma_t \label{eq:objP}\\
\text { s. t. } & \gamma_{i,k} \geq \Gamma, i=1, \ldots, N_{\mathrm{Ue}}, k=1, \ldots, N_{\mathrm{Tx}}\label{eq:gammai}\\
& \sum_{i=1}^{N_{\mathrm{Ue}}}||\mathbf{w}_{i,k}||^2\!+\!||\mathbf{w}_{t,k}||^2\!\leq \!P_T, k=1,...,N_{\mathrm{Tx}},\label{eq:pt}
\end{align}
\end{subequations}
where $\Gamma$ is the communication SINR threshold, and $P_T$ is the transmit power limit for the transmitter APs. To solve $\mathcal{P}$, an iterative process is employed. At first, the UEs receive beamformers $\mathbf{u}_i$, $i$$=$$1,...,N_{\mathrm{Ue}}$ are randomly chosen and fixed. Nonetheless, note that the optimization problem in $\mathcal{P}$ is still not convex due to the non-concave objective function and non-convex constraint in \eqref{eq:gammai}. To handle this, \eqref{eq:objP} can be split into two sub-problems as
\begin{subequations}
\begin{align}
\mathcal{P}1\!:\! \max_{\mathbf{W}} &\zeta\!\Bigg(\!\sum_{n=1}^{N\!-\!1}\!\!\mathbf{s}^H\![n]\!\Bigg(\!\sum_{r=1}^{N_{\mathrm{Rx}}}\! \sum_{k=1}^{N_{\mathrm{Tx}}}\! \varphi_{r,k,k}\mathbf{W}_k^H\mathbf{A}_{k,k}\mathbf{W}_k\!\Bigg)\!\mathbf{s}[n]\!\!\Bigg)\label{eq:p1}\\
\text { s. t. } & \eqref{eq:gammai}, \eqref{eq:pt},\nonumber 
\end{align}
\end{subequations}
and, 
\begin{subequations}
\begin{align}
\mathcal{P}2\!: \max_{\mathbf{W}} &\zeta\!\Bigg(\!\sum_{n=1}^{N-1}\!\!\mathbf{s}^H\![n]\!\Bigg(\!\sum_{r=1}^{N_{\mathrm{Rx}}}\! \sum_{k=1}^{N_{\mathrm{Tx}}}\! \sum_{j=1 \atop j \neq k}^{N_{\mathrm{Tx}}}\! \varphi_{r,k,j}\mathbf{W}_k^H\mathbf{A}_{k,j}\mathbf{W}_j\!\Bigg) \mathbf{s}[n]\!\Bigg)\label{eq:p2}\\
\text { s. t. } & \eqref{eq:gammai},\eqref{eq:pt}.\nonumber 
\end{align}
\end{subequations}
Next, as in~\cite{art:paperRFVLC}, we consider the first-order Taylor approximation to linearize \eqref{eq:p1} and \eqref{eq:gammai}, and employ the constrained convex-concave procedure (CCCP) to iteratively solve the approximated convex problem until a convergence criterion is attained. Thus, at the $p$th iteration, $\mathcal{P}1$ is solved as
\begin{subequations}
\begin{align}
\mathcal{P}1'&\!:\! \max_{\mathbf{W}, \tau_{i,k}, \mu_{i,k}}\!\! \zeta\!\Bigg(\!\sum_{n=1}^{N\!-\!1}\!\!\mathbf{s}^H\![n]\!\Bigg(\!\sum_{r=1}^{N_{\mathrm{Rx}}}\! \sum_{k=1}^{N_{\mathrm{Tx}}}\! \varphi_{r,k,k}\!\Bigg(\!\!2\left[\!\mathbf{W}_k^{(p\!-\!1)}\!\right]^H\!\!\!\mathbf{A}_{k,k}\nonumber\\
\!\times&\!\!\Bigg(\!\mathbf{W}_k\!-\!\mathbf{W}_k^{(p\!-\!1)}\!\!\Bigg)\!\!+\!\!
\left[\!\mathbf{W}_k^{(p\!-\!1)}\!\right]^H\!\!\!\mathbf{A}_{k,k}\mathbf{W}_k^{(p\!-\!1)}\!\Bigg)\!\!\!\Bigg)\!\mathbf{s}[n]\!\Bigg)\label{eq:p1linha}\\
\text { s. t. }\!&|\mathbf{u}_i^H\mathbf{H}_{i,k}\mathbf{w}_{i,k}^{(p-1)}|^2\!+\!2\!\left[\!\mathbf{w}_{i,k}^{(p-1)}\!\right]^H\!\!\mathbf{H}_{i,k}^H\mathbf{u}_i\mathbf{u}_i^H\mathbf{H}_{i,k}\nonumber\\
&\times\!\left(\!\mathbf{w}_{i,k}\!-\!\mathbf{w}_{i,k}^{(p-1)}\!\right)\!\geq \tau_{i,k}, \forall i, \forall k\label{eq:SINR1}\\
&\sum_{l=1 \atop l \neq i}^{N_{\mathrm{Ue}}}\!|\mathbf{u}_i^H\mathbf{H}_{i,k}\mathbf{w}_{l,k}|^2\!+\!|\mathbf{u}_i^H\mathbf{H}_{i,k}\mathbf{w}_{t,k}|^2\!+\!\sigma^2_i||\mathbf{u}_i||^2\!\leq\!\mu_{i,k}\label{eq:SINR2}\\
&\tau_{i,k} \geq \Gamma \mu_{i,k}, \forall i, \forall k,\label{eq:slack}\\
&\eqref{eq:pt},\nonumber 
\end{align}
\end{subequations}
where $\tau$ and $\mu$ are slack variables. $\mathcal{P}1'$ is a concave problem. 
$\mathcal{P}2$, on the other hand, is approximated to the following
\begin{subequations}
\begin{align}
\mathcal{P}2': \max_{\mathbf{W}, \tau_{i,k},\mu_{i,k}} &\zeta\!\Bigg(\!\sum_{n=1}^{N-1}\!\!\mathbf{s}^H\![n]\!\Bigg(\!\sum_{r=1}^{N_{\mathrm{Rx}}}\! \sum_{k=1}^{N_{\mathrm{Tx}}}\! \sum_{j=1 \atop j \neq k}^{N_{\mathrm{Tx}}}\! \varphi_{r,k,j}\mathbf{W}_k^H\mathbf{A}_{k,j}\nonumber\\&\times\mathbf{W}_j^{(p-1)}\Bigg)\mathbf{s}[n]\Bigg)\label{eq:p21}\\
\text { s. t. } & \eqref{eq:SINR1}, \eqref{eq:SINR2},\eqref{eq:slack}, \eqref{eq:pt}.\nonumber 
\end{align}
\end{subequations}
which is a concave problem. Accordingly, both optimization can be solved with convex toolboxes such as CVX. After obtaining the optimal $p$th $\mathbf{W}^*$, given by the summation of the solutions of $\mathcal{P}1'$ and $\mathcal{P}2'$, the transmit precoder matrices are fixed and, for all $i$, $\mathbf{u}_i$ is updated via the MMSE receiver as~\cite{7405344}
\begin{align}
         \mathbf{u}_i\!&=\!\!\sum_{k=1}^{N_{\mathrm{Tx}}}\!\!\left(\!\mathbf{H}_{i,k}\!\!\left(\!\sum_{l=1 \atop l \neq i}^{N_{\mathrm{Ue}}}\mathbf{w}_{l,k}\mathbf{w}_{l,k}\!\!\right)\!\mathbf{H}_{i,k}^H\!+\!\sigma_i^2\mathbf{I}_{M_{\mathrm{Ue}}}\!\!\right)^{\!-\!1}\!\!\!\mathbf{H}_{i,k} \mathbf{w}_{i,k}. \label{eq:ui}
     \end{align}
     
Accordingly, the algorithm to obtain the transmit and receive precoders is summarized in Algorithm 1. 
\begin{algorithm}\label{alg:CCCP}
\caption{Precoders Iterative Algorithm}
\begin{algorithmic}[1]
\footnotesize
       \State Choose error tolerances $\epsilon$ , maximum number of iterations $i_{max}$, and feasible initial points $\mathbf{W}^{(0)}$, and $\mathbf{u}_i(0)$ $\forall i$.
       \State $p \gets 0$
       \Repeat
       \State Solve \eqref{eq:p1linha} and \eqref{eq:p21} to obtain $\mathbf{W}^{*(p)}$ using $\mathbf{W}^{*(p-1)}$ and $\mathbf{u}_{i}^{(p-1)}$ from the previous iteration.
       \State Solve \eqref{eq:ui} using $\mathbf{W}^{*(p)}$ to attain $\mathbf{u}_{i}^{(p)}$.
       \State $p++$
       \Until{$\frac{||\mathbf{W}^{*(p)}-\mathbf{W}^{*(p-1)}||}{\mathbf{W}^{*(p)}}$$\leq$$\epsilon$ and $\frac{||\mathbf{u}_i^{*(p)}-\mathbf{u}_i^{*(p-1)}||}{\mathbf{u}_i^{*(p)}}$$\leq$$\epsilon$ or $p$$\geq$$i_{max}$}
      \State \Return $\mathbf{W}^*$, $\mathbf{u}_i^*$ $\forall i$.  
\end{algorithmic}
\end{algorithm}
\vspace{-0.47cm}
\section{Adversary Model}
Assuming that the UEs know that there is a present target in the network, one of them may act as an adversary. Given its received signal, the UE could try to infer the position of the target by recreating a replica of the transmit beampattern created by the transmitter APs given by
\begin{align}\label{eq:beampat}
    \mathbf{B}_k = \mathbf{a}^H(\theta_{z})\mathbf{R}_{\mathbf{x}_k}\mathbf{a}(\theta_{z}), \forall k,
\end{align}
where $\{\theta_z\}^{Z}_{z=1}$ are sampled angle grids, and $\mathbf{R}_{\mathbf{x}_k}$$=$$\mathbf{x}_k\mathbf{x}_k^H$ is the covariance matrix of the transmitted signal. To accomplish this, it is considered that the adversary has knowledge of its receive beamforming vector. Also, it is feasible to assume that the adversary knows the position of the $N_{\mathrm{Tx}}$ transmitter APs and can create a search area based on that information. Under these considerations, the adversary estimation process starts with the inference of the interference signal from its received signal, which is expressed as in \eqref{eq:signali}, is equal to
\begin{align}
     \tilde{\mathbf{y}}_{a,k} = \mathbf{H}_{a,k}\underbrace{\left(\sum_{l=1 \atop l \neq a}^{N_{\mathrm{Ue}}}\!\mathbf{w}_{l,k}s_l[n]+\mathbf{w}_{t,k}s_t[n]\right)}_{\tilde{\mathbf{x}}_k}+\mathbf{n}_a.
 \end{align}
 We further assume that the true $\mathbf{H}_{a,k}$ is not available at the adversary, but can be estimated as $\mathbf{H}_{a,k}\!=\!\Hat{\mathbf{H}}_{a,k}\!+\!\varepsilon$, where $\Hat{\mathbf{H}}_{a,k}$ is its channel estimate, which can be attained via least squares estimation with pilot signals as
\begin{align}
    \Hat{\mathbf{H}}_{a,k} = \mathbf{y}_{a,k}\mathbf{x}_k^H(\mathbf{x}_k\mathbf{x}_k^H)^{-1},
\end{align}
and $\varepsilon$ represents the channel estimation error, modeled as $\mathcal{N}(0,\sigma_\varepsilon^2 \mathbf{I})$. Thus, $\mathbf{H}_{a,k}$$\sim$$\mathcal{N}(\Hat{\mathbf{H}}_{a,k},\sigma_\varepsilon^2 \mathbf{I})$.

Then, considering  $\tilde{\mathbf{y}}_{a,k}$ as an observable variable, $\mathbf{H}_{a,k}$ as a latent variable, and   $\tilde{\mathbf{x}}_k$ as the unknown parameter, an iterative method to compute the maximum log-likelihood as the EM can be employed to estimate $\tilde{\mathbf{x}}_k$. The EM algorithm is a general technique used to find the maximum likelihood estimates of parameters when part of the data is missing or for latent variable models. The algorithm is divided into two steps, the E-step and the M-step. In the former, the expected value of the log-likelihood function is computed, as well as a current estimate of the unknown parameter. In the latter, the unknown parameter is attained from the maximization of the previously obtained expected value~\cite{art:em}.
\subsection{Expectation-Maximization Algorithm}
As the goal is to maximize the log-likelihood function, we start by defining it as
\begin{align}\label{eq:likelihood}
    \mathcal{L}\!=\!\log \left(p_{\tilde{\mathbf{y}}_{a,k}}\!({\tilde{\mathbf{x}}_k})\right)\!=\!\log\!\!\left(\int\!\! p_{\tilde{\mathbf{y}}_{a,k},\mathbf{H}_{a,k}}\!(\tilde{\mathbf{x}}_{k}) d\mathbf{H}_{a,k}\right), 
\end{align}
where $p_{\tilde{\mathbf{y}}_{a,k}}$ is the probability density function (PDF) of the variable $\tilde{\mathbf{y}}_{a,k}$, and $p_{\tilde{\mathbf{y}}_{a,k},\mathbf{H}_{a,k}}\!(\tilde{\mathbf{x}}_{k})$ stands for the joint PDF of $\mathbf{y}_{a,k}$ and $\mathbf{H}_{a,k}$, and can be rewritten as
\begin{align}
   p_{\tilde{\mathbf{y}}_{a,k},\mathbf{H}_{a,k}}\!(\tilde{\mathbf{x}}_{k}) = p_{\mathbf{H}_{a,k}}\!(\mathbf{H}_{a,k})p_{\tilde{\mathbf{y}}_{a,k}|\mathbf{H}_{a,k}}\!(\tilde{\mathbf{x}}_{k}),
\end{align}
where $p_{\tilde{\mathbf{y}}_{a,k}|\mathbf{H}_{a,k}}\!(\tilde{\mathbf{x}}_{k})$ is the conditional pdf of $\tilde{\mathbf{y}}_{a,k}$ and $\mathbf{H}_{a,k}$. As in~\cite{9882359}, to overcome the sensitivity of the Gaussian distribution to outliers, the heavy-tailed Student`s $t$ distribution is considered to describe $p_{\tilde{\mathbf{y}}_{a,k}|\mathbf{H}_{a,k}}\!(\tilde{\mathbf{x}}_{k})$, that is,
\begin{align}\label{eq:student}
    p_{\tilde{\mathbf{y}}_{a,k}|\mathbf{H}_{a,k}}\!(\tilde{\mathbf{x}}_{k})\!&\sim\!lst\left(\mathbf{H}_{a,k}\tilde{\mathbf{x}}_k, \sigma_{a}^2 \mathbf{I}, v\right),\nonumber\\
    &=\!\int_{0}^{\infty}\!\!\mathcal{N}\!\left(\!\mathbf{H}_{a,k}\tilde{\mathbf{x}}_{k}, \frac{\sigma_{a}^2 \mathbf{I}}{u}\!\right)\!\mathcal{G}\!\left(\!\frac{v}{2}, \frac{v}{2}\!\right)\!d u,
\end{align}
where $v$ is the degree of freedom, $u$ is an auxiliary hidden variable, and $\mathcal{G}$ is the gamma function. Thus, the latent variable is redefined as $\mathbf{h}_{a,k}=\left\{\mathbf{H}_{a,k}, u \right\}$, and the model parameters are given by $\mathbf{\vartheta}=\left\{\tilde{\mathbf{x}}_{k}, v\right\}$. Next, based on~\cite{art:em}, by introducing an arbitrary distribution $q(\mathbf{h}_{a,k})$ over the latent variable $\mathbf{h}_{a,k}$, $\mathcal{L}$ can be decomposed in two terms as
\begin{align}
	\mathcal{L}  \!=\!&\int q(\mathbf{h}_{a,k}) \log \left(p_{\tilde{\mathbf{y}}_{a,k},\mathbf{h}_{a,k}}\left(\mathbf{\vartheta}\right)\right) d \mathbf{h}_{a,k}, \nonumber\\
	 \!=\!&\int q(\mathbf{h}_{a,k}) \log\left( \frac{p_{\tilde{\mathbf{y}}_{a,k},\mathbf{h}_{a,k}}\left(\mathbf{\vartheta}\right)}{q(\mathbf{h}_{a,k})}\frac{q(\mathbf{h}_{a,k})}{p_{\mathbf{h}_{a,k}|\tilde{\mathbf{y}}_{a,k}}\left(\mathbf{\vartheta}\right)}\right) d \mathbf{h}_{a,k}, \nonumber\\
	\!=\!&\underbrace{\int q(\mathbf{h}_{a,k}) \log \left(\frac{p_{\tilde{\mathbf{y}}_{a,k},\mathbf{h}_{a,k}}\left(\mathbf{\vartheta}\right)}{q(\mathbf{h}_{a,k})}\right)d \mathbf{h}_{a,k}}_{\mathcal{F}_{\tilde{\mathbf{y}}_{a,k}}\!\left(q, \mathbf{\vartheta}\right)}+\mathrm{KLD},
\end{align}


where $\mathrm{KLD}$ is given by
\begin{align}
	\mathrm{KLD}\!
 =\!\!\!\int\!\! q(\mathbf{h}_{a,k})\! \log\!\left(\! \frac{q(\mathbf{h}_{a,k})}{p_{\mathbf{h}_{a,k}|\tilde{\mathbf{y}}_{a,k}}\left(\mathbf{\vartheta}\right)}\!\right)\!d \mathbf{h}_{a,k},
\end{align}
which is the Kullback-Leiber divergence, and has the property of always being greater or equal to zero. Hence, the E-step and the M-step of the EM algorithm can be described only in terms of $\mathcal{F}_{\tilde{\mathbf{y}}_{a,k}}\!\left(q, \mathbf{\vartheta}\right)$ as presented next.
 \subsubsection{E-step} 
 $\mathcal{F}$ is maximized in terms of $q(\mathbf{h}_{a,k})$ for a fixed $\mathbf{\vartheta}$. Thus, after every $(a\!+\!1)$th iteration, $q(\mathbf{h}_{a,k})$ is updated as
\begin{align}
	q^{a+1}(\mathbf{h}_{a,k})=\arg \max _{q(\mathbf{h}_{a,k})} \mathcal{F}_{\tilde{\mathbf{y}}_{a,k}}\!\left(q, \mathbf{\vartheta}^a\right) .
\end{align}
 Given that $\mathbf{H}_{a,k}$ and $u$ are independent, $q$ can be factorized as $q(\mathbf{h}_{a,k})$ = $q(\mathbf{H}_{a,k})q(u)$. And based on~\cite{9882359}, $q(\mathbf{H}_{a,k})$ and $q(u)$ are optimal as
\begin{align}
	&q(\mathbf{H}_{a,k})\!\!\propto p_{\mathbf{H}_{a,k}}\!(\mathbf{H}_{a,k})\!\exp\!\left\{\!\mathbb{E}_{u}\!\left[\!\log\!\mathcal{N}\!\left(\!\mathbf{H}_{a,k}\tilde{\mathbf{x}}_{k}, \frac{\sigma_{a}^2 \mathbf{I}}{u}\!\right)\!\right]\!\right\},\label{eq:qhak} \\
	&q\!\left(u\right)\!\!\propto\mathcal{G}\!\!\left(\!\frac{v}{2}, \frac{v}{2}\right)\!\exp\!\left\{\!\mathbb{E}_{\mathbf{H}_{a,k}}\!\!\left[\!\log \mathcal{N}\!\left(\!\mathbf{H}_{a,k}\tilde{\mathbf{x}}_{k}, \!\frac{\sigma_{a}^2 \mathbf{I}}{u}\!\right)\!\right]\!\right\}\!.\label{eq:qu}
\end{align}
Accordingly, the distribution in \eqref{eq:qhak} have the mean and covariance given, respectively, by
\begin{align}
	\mathbb{E}_{\mathbf{H}_{a,k}}[q(\mathbf{H}_{a,k})] & \!=\!\!\left[\boldsymbol{\Omega}_{\mathbf{H}}\!\left(\frac{1}{\sigma_{\epsilon}^{2}} \Hat{\mathbf{H}}_{a,k}^{H}\!+\!\tilde{\mathbf{x}}_{k} \tilde{\mathbf{y}}_{a,k}^H \frac{\mathbb{E}_{u}\left[u\right]}{\sigma_{a}^2}\right)\!\right]^{H}\!,\label{eq:media}\\
	\boldsymbol{\Omega}_{\mathbf{H}_{a,k}} & =\left[\frac{1}{\sigma_{\epsilon}^{2}} \mathbf{I}+\tilde{\mathbf{x}}_{k}\tilde{\mathbf{x}}_{k}^{H} \frac{\mathbb{E}_{u}\left[u\right]}{\sigma_{a}^2}\right]^{-1}.\label{eq:rowcov}
\end{align}  
$q\left(u\right)$ on the other hand, can be approximated to $\mathcal{G}\!\left(\!\frac{v+2 N}{2},\!\frac{v+C}{2}\!\right)$, with
\begin{align}
	C\!=\!&\Bigg(\!\!\!\left(\tilde{\mathbf{y}}_{a,k}\!-\!\mathbb{E}_{\mathbf{H}_{a,k}}[q(\mathbf{H}_{a,k})] \tilde{\mathbf{x}}_{k}\right)^{H}\!\!\left(\tilde{\mathbf{y}}_{a,k}\!-\!\mathbb{E}_{\mathbf{H}_{a,k}}[q(\mathbf{H}_{a,k})] \tilde{\mathbf{x}}_{k}\right)\nonumber\\&+\tilde{\mathbf{x}}_{k}^{H}\left[\boldsymbol{\Omega}_{\mathbf{H}_{a,k}} M_{\mathrm{AP}}\right] \tilde{\mathbf{x}}_{k}\!\!\Bigg)\frac{1}{\sigma_{a}^2}.
\end{align}
\subsubsection{M-step}
For the variational $\mathrm{M}$-step, $\mathcal{F}$ is maximized in terms of $\mathbf{\vartheta}$ with fixed $q(\mathbf{h}_{a,k})$. Again, after every $(a\!+\!1)$th iteration, $\mathbf{\vartheta}^{a+1}$ is obtained by
\begin{align}
	\mathbf{\vartheta}^{a+1}=\arg \max _{\mathbf{\vartheta}} \mathcal{F}_{\tilde{\mathbf{y}}_{a,k}}\!\left(q^{a+1}, \mathbf{\vartheta}\right).
\end{align}
\vspace{0.5cm}
 Given \eqref{eq:media} and \eqref{eq:rowcov}, and after some mathematical manipulations,  $\mathcal{F}_{\tilde{\mathbf{y}}_{a,k}}\!\left(q^{a+1}, \mathbf{\vartheta}\right)$ can be written as $\sigma_{a}^2C.$
Thus, the estimated transmitted interference signal, $\Hat{\mathbf{x}}_{k}$  can be recovered by minimizing $\mathcal{F}_{\tilde{\mathbf{y}}_{a,k}}\!\left(q^{a+1}, \mathbf{\vartheta}\right)$. 
However, to facilitate the analysis, a Choleski decomposition may be applied on $\boldsymbol{\Omega}_{\mathbf{H}_{a,k}}  M_{\mathrm{AP}}=\mathbf{V}_{\mathbf{H}}^{\top} \mathbf{V}_{\mathbf{H}}$, and $\Hat{\mathbf{x}}_{k}$ is obtained via the following optimization
\begin{align}\label{eq:optimalx}
	\Hat{\mathbf{x}}_{k} & \!=\!\min\!\left(\left\|\tilde{\mathbf{y}}_{a,k}\!-\!\mathbb{E}_{\mathbf{H}_{a,k}}[q(\mathbf{H}_{a,k})] \tilde{\mathbf{x}}_{k}\right\|^{2}\!\!+\!\left\|\mathbf{V}_{\mathbf{H}} \tilde{\mathbf{x}}_{k}\right\|^{2}\right).
\end{align}

The EM implementation is described in Algorithm 2.
\begin{algorithm}\label{alg:coalitional}
\caption{EM Algorithm}
\begin{algorithmic}[1]
\footnotesize		 \Initialize{$\mathbb{E}_{\mathbf{H}_{a,k}}^{(0)}[q(\mathbf{H}_{a,k})]\gets \hat{\mathbf{H}}_{a,k}$; $\Hat{\mathbf{x}}_{k}^{(0)}\gets \Hat{\mathbf{x}}_{k}^{\mathrm{ZF}}\ $; $a \gets 0$}

		\While{convergence = 0}
        \State Calculate $q^{(a+1)}(u)$ and $\mathbb{E}_u[u]^{(a+1)}$.
		\State Update $\mathbb{E}_{\mathbf{H}_{a,k}}^{(a+1)}$ according to \eqref{eq:media}, and $\boldsymbol{\Omega}^{(a+1)}_{\mathbf{H}}$ as in \eqref{eq:rowcov}.
		\State Do the Cholesky decomposition:
			$\boldsymbol{\Omega}_{\mathbf{H}}^{\mathrm{(a+1)}}\!M_{\mathrm{AP}}$$=$$\mathbf{V}_{\mathbf{H}}^{(a +1)T} \mathbf{V}_{\mathbf{H}}^{(a+1)}$.
	\State Solve \eqref{eq:optimalx} to achieve $\Hat{\mathbf{x}}_k^{(a+1)}$ given $\mathbf{V}_{\mathbf{H}}^{(a+1)}$ and $\mathbb{E}_{\mathbf{H}_{a,k}}^{(a+1)}$.
		\If{$||\hat{\mathbf{x}}_k^{a}-\hat{\mathbf{x}}_k^{a-1}|| \leq cov$}
		\State convergence = 1
		\Else
		\State $a \gets a+1$;
		\State Return to Step 3.
		\EndIf
		\EndWhile
		\State \Return $\hat{\mathbf{x}}_k^{a}$.
\end{algorithmic}
\end{algorithm}

In Algorithm 2, $\hat{\mathbf{x}}_k^{\mathrm{ZF}}$ is computed via a zero-forcing (ZF) detector as 
\begin{align}
    \hat{\mathbf{x}}_k^{\mathrm{ZF}} = (\hat{\mathbf{H}}_{a,k}^H\hat{\mathbf{H}}_{a,k})^{-1}\hat{\mathbf{H}}_{a,k}^H\mathbf{y}_{a,k}.
\end{align}

Accordingly, after $\hat{\mathbf{x}}_k$ is obtained, its covariance matrix is computed as $\mathbf{R}_{\hat{\mathbf{x}}_k}$$=$$\hat{\mathbf{x}}_k\hat{\mathbf{x}}_k^H$, and the estimated beampattern for the $k$th transmitter AP, $\hat{\mathbf{B}}_k$, is designed as in~\eqref{eq:beampat}.

\begin{figure}[t]
    \centering
    \includegraphics[scale=0.14]{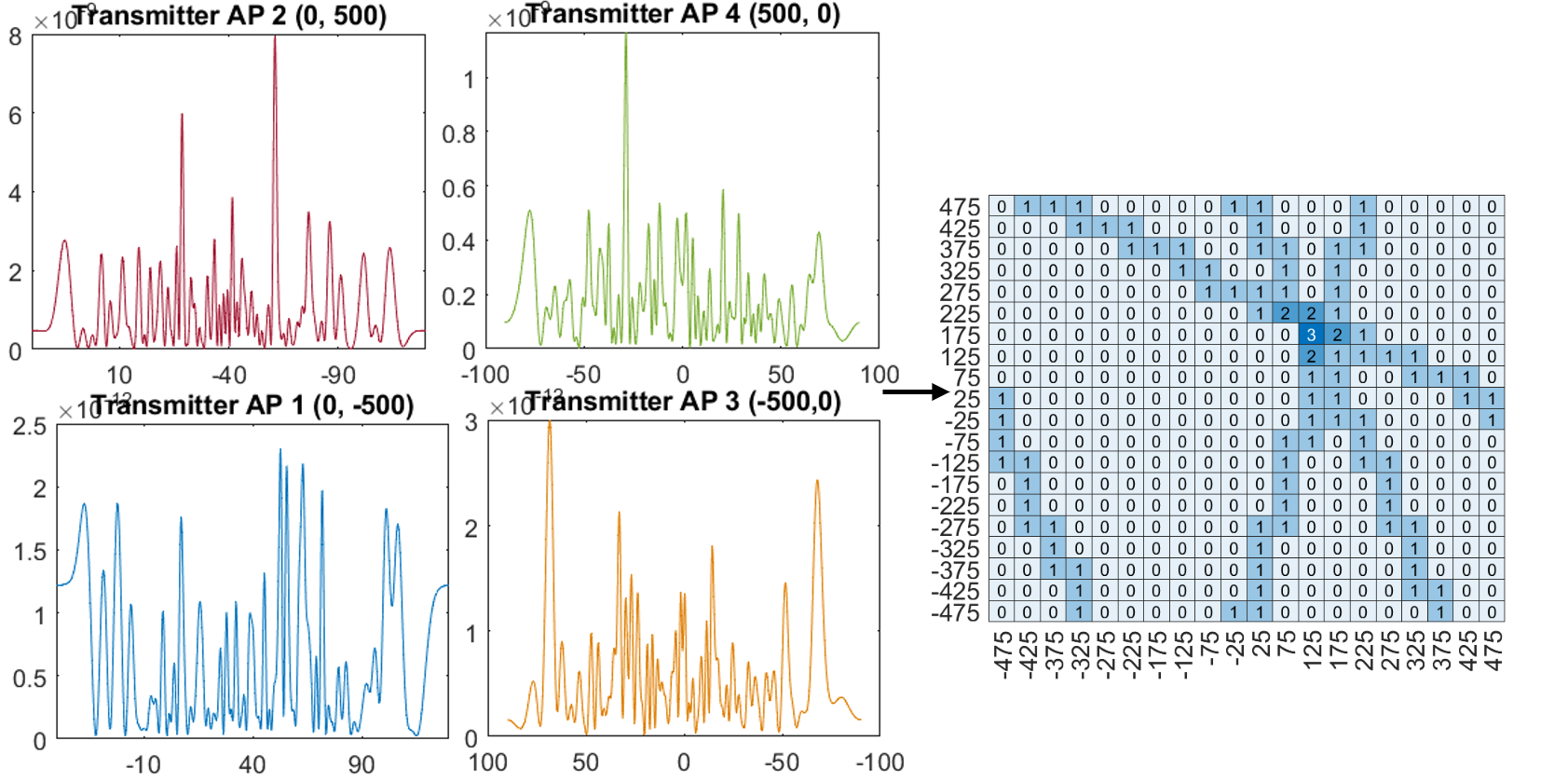}
    \caption{Schematic of the process to compute the location of the target. On the left side, the estimated transmit beampatterns for each transmit AP are shown, and on the right side is a heat map of the search area for the computed direction lines from the transmit APs.}
    \label{fig:EM}
    \vspace{-0.52cm}
\end{figure}

As illustrated in Fig.~\ref{fig:EM}, the adversary considers the maximum angle from $\hat{\mathbf{B}}_k$ as the estimated direction of the beam of AP $k$ to the target and defines a line of search inscribed in a square search area, having as many lines as number of transmitter APs. Next, the adversary divides the search area in cells of same size and verifies which of them are crossed by the computed direction lines. At the end, the cell crossed by the highest number of lines is set as the estimated position for the target. In case more than one cell contains the maximum number of crossing lines, the estimated position is selected randomly among the cells with maximum value.
Thus, the probability of detection for an observation $q$ is given by
\begin{align}
    P_{D,q} = \begin{cases}
      1 & \text{if $\hat{Pos_{t,q}}=Pos_t$}\\
      0 & \text{otherwise},
    \end{cases}       
\end{align}
where $Pos_t$ and $\hat{Pos}_t$ are, respectively, the true and estimated position of the target. Hence the probability of detection for the total number of observations $Q$ is given by $P_D$$=$$\sum_{q=1}^{Q}P_{D,q}/Q$. 
\section{Numerical results and discussions}
For the numerical results, the search area is assumed to have 1000m $\times$ 1000 m, with cells of 50m $\times$ 50m and the coordinate $(0,0)$ positioned at the center of the search area. It is considered $N_{\mathrm{Tx}}$$=$$8$ transmitter APs serving $N_{\mathrm{Ue}}$$=$$4$ UEs with $M_{\mathrm{Ue}}$$=$$8$ antennas each. It is also assumed $N_{\mathrm{Rx}}$$=$$4$ sensing receiver APs, and both transmitter and receiver APS are equipped with $M_{\mathrm{AP}}$$=$$64$ antennas. The noise variance for each user, $\sigma_i^2$ , the communication SINR threshold, $\Gamma$, and the maximum transmit power per AP, $P_T$ are respectively set as -94 dBm, 3 dBm and 50 dBm. The carrier frequency is $\lambda_c$$=$$1.9$ GHz, and the variance of the RCS is $\sigma_{r,k}^2$$=$$10$ dBsm. In addition,  the average channel gains of the communication links are assumed to be determined by the pathloss, i.e., $\Omega_{i,k}$$=$$d_{i,k}^{-\varphi}$, where $d_{i,k}$ is the distance between the $k$th transmitter AP and UE $i$, and the path-loss exponent is set as $\varphi$$=$$3$. For algorithm 1,  the error tolerance $\epsilon$ is set as 0.1 and $i_{max}$$=$$10$ proved to render a sufficient number of iterations. Moreover, $\mathbf{W}^{(0)}$ and $\mathbf{u}_i(0)$ are chosen randomly. For algorithm 2, the variance of the channel estimation error is $\sigma_{\varepsilon}^2$$=$$10$ dBm and $cov$$=$$10^{-5}$. $P_D$ is computed over $Q$$=$$100$ channel realizations, and unless specified otherwise the coordinates of the target is $(-75,75)$, the receiver APs are at $(250, 250)$, $(-250,-250)$, $(-250,250)$ and $(250,-250)$. The UEs are positioned at $(300, 300)$, $(-300,-300)$, $(-300,300)$ and $(300,-300)$, where the first one is considered as the adversary. Finally, the eight transmitter APS are at $(-500, -500)$, $(500,500)$, $(-500,500)$, $(500,-500)$, $(0,-500)$, $(0,500)$, $(-500,0)$ and $(500,0)$. 


\begin{figure*}[!tbp]
  \begin{subfigure}[t]{0.3\textwidth}
    \includegraphics[width=\textwidth]{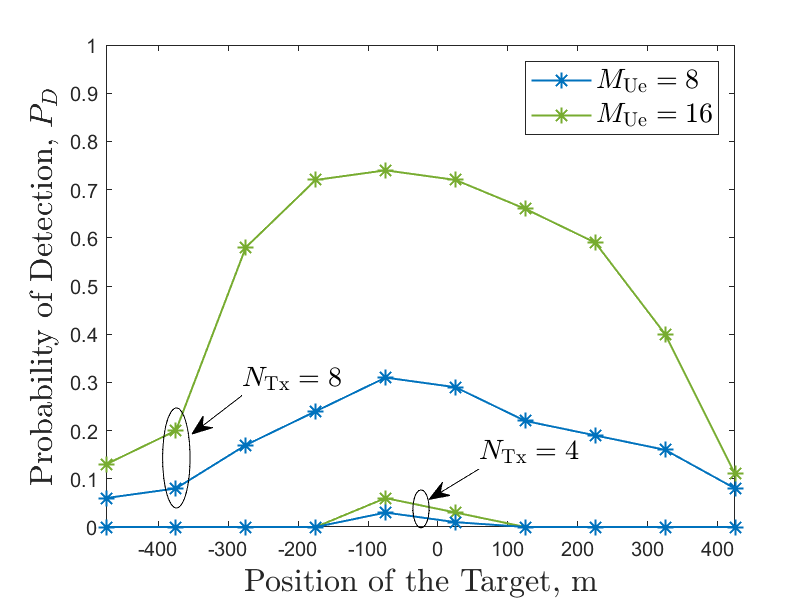}
    \caption{$M_{\mathrm{Ue}}$$=$$8, 16$, $N_{\mathrm{Tx}}$$=$$4, 8$.}
    \label{fig:postarget}
  \end{subfigure}
  \hfill 
  \begin{subfigure}[t]{0.3\textwidth}
    \includegraphics[width=\textwidth]{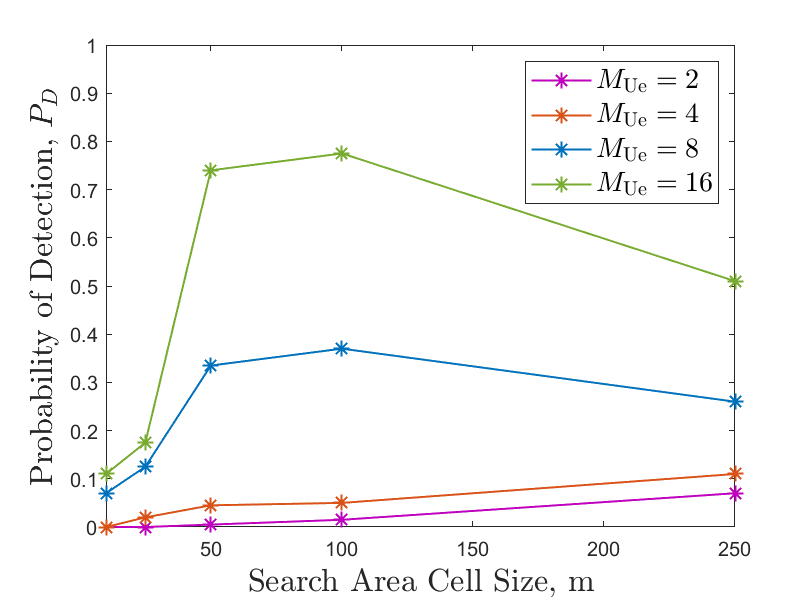}
    \caption{$M_{\mathrm{Ue}}$$=$$2,4, 8, 16$.}
    \label{fig:cellsize}
  \end{subfigure}
  \hfill 
  \begin{subfigure}[t]{0.3\textwidth}
    \includegraphics[width=\textwidth]{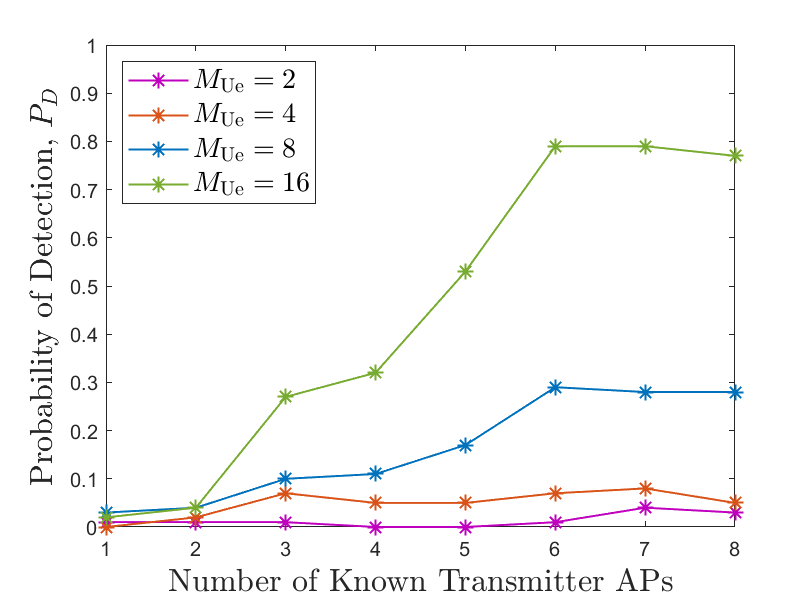}
    \caption{$M_{\mathrm{Ue}}$$=$$2,4, 8, 16$.}
    \label{fig:knownaps}
  \end{subfigure}
  \caption{Probability of detection versus (a) the position of the target, (b) the search area cell size, and (c) number of known transmitter APs.}
\vspace{-0.52cm}
\end{figure*}

Fig.~\ref{fig:postarget} illustrates $P_D$ vs. the position of the target for $M_{\mathrm{Ue}}$$=$$8, 16$ and $N_{\mathrm{Tx}}$$=$$4, 8$. For this figure, the target's position is varied along the main diagonal of the search area, with the label of Fig.~\ref{fig:postarget} indicating the variation along the x-axis. Also, the coordinates of transmitter APS for the case with $N_{\mathrm{Tx}}$$=$$4$ are $(0,-500)$, $(0,500)$, $(-500,0)$ and $(500,0)$. Note that larger number of antennas at UEs increases the probability of detection by the adversary, as expected. Although the increase is noticeably more pronounced when the number of transmitter APs is also larger, as it increases the possibility that a bigger number of estimated direction lines converges to the cell that contains the target. Moreover, notice that the maximum $P_D$ for all cases occurs when the target is positioned closer to the center of the search area, which is also the closer point to the adversary location. In that case, the interference from the sensing signal at the adversary is higher, thus allowing for a better estimation from the EM algorithm. For the worst case, with $N_{\mathrm{Tx}}$$=$$8$ and $M_{\mathrm{Ue}}$$=$$16$, it leads to a $P_D$ higher than 0.7 which highlights the risk for the target's location privacy.


In Fig.~\ref{fig:cellsize}, $P_D$ is evaluated for different sizes of cells and different values of the number of antennas at the UEs, $M_{\mathrm{Ue}}$$=$$2,4, 8, 16$. Validating the results observed in the previous figure, by increasing the number of antennas at the UEs, $P_D$ also increases. Note that for cases $M_{\mathrm{Ue}}$$=$$8$ and $M_{\mathrm{Ue}}$$=$$16$, there is a decrease on $P_D$ when the cell size is bigger than 100 m. It can be explained by the fact that there are fewer cells within the search area as the cell size increases, thus more cells might contain the maximum number of crossing lines, which reduces the probability that the chosen estimate position be the correct one. Even though, the $P_D$ remains relatively high, with the adversary identifying the correct cell in more than 50\% of the cases.


Fig.~\ref{fig:knownaps} illustrates $P_D$ vs. the number of known transmitter APs position from the adversary for $M_{\mathrm{Ue}}$$=$$2,4, 8, 16$. For this figure, we consider that the adversary receives data from all transmitter APs, but might not be capable to identify the location of all them. Note that, for $M_{\mathrm{Ue}}$$=$$8, 16$, with the knowledge of 6 transmitter APs, the adversary is already capable to achieve the maximum $P_D$ for both values of $M_{\mathrm{Ue}}$. Also, even with the position information of fewer APs, the adversary can still correctly detect the target in several opportunities. 

\section{Conclusions}
In this paper, the capability of an UE, acting as adversary, to infer the position of a target in a multi-static sensing-centric ISAC system was investigated. For the adversary model, an EM algorithm was proposed to obtain an estimation of the transmitted signal, which was used to create replicas of the transmit beampattern of each transmitter AP. From the results, we showed that as more APs are deployed than transmitters, the adversary is capable to successfully identify the target location in a reasonable number of cases. Different from the mono-static scenario, where the adversary is capable to infer the angular position of the target, the multi-static sensing-centric scenario showed to be more critical in terms of privacy, since the adversary is capable of, with some accuracy, identify the exact location of the target. Such threats becomes even more critical as the network becomes denser.  

\section*{Acknowledgement}
This work has been supported by Academy of Finland, 6G Flagship program (Grant 346208) and FAITH Project (Grant 334280). Codes to reproduce results are available at: https://github.com/isabella-gomes/Globecom2023
\bibliographystyle{IEEEtran}
\bibliography{references.bib}

\begin{thebibliography}{10}
\providecommand{\url}[1]{#1}
\csname url@samestyle\endcsname
\providecommand{\newblock}{\relax}
\providecommand{\bibinfo}[2]{#2}
\providecommand{\BIBentrySTDinterwordspacing}{\spaceskip=0pt\relax}
\providecommand{\BIBentryALTinterwordstretchfactor}{4}
\providecommand{\BIBentryALTinterwordspacing}{\spaceskip=\fontdimen2\font plus
\BIBentryALTinterwordstretchfactor\fontdimen3\font minus
  \fontdimen4\font\relax}
\providecommand{\BIBforeignlanguage}[2]{{%
\expandafter\ifx\csname l@#1\endcsname\relax
\typeout{** WARNING: IEEEtran.bst: No hyphenation pattern has been}%
\typeout{** loaded for the language `#1'. Using the pattern for}%
\typeout{** the default language instead.}%
\else
\language=\csname l@#1\endcsname
\fi
#2}}
\providecommand{\BIBdecl}{\relax}
\BIBdecl

\bibitem{9737357}
F.~Liu and {Et. al.}, ``Integrated sensing and communications: Toward
  dual-functional wireless networks for {6G} and beyond,'' \emph{IEEE Journal
  on Sel. Areas in Commun.}, vol.~40, no.~6, pp. 1728--1767, 2022.

\bibitem{Art:Emil}
Z.~Behdad and {Et. al.}, ``Multi-static target detection and power allocation
  for integrated sensing and communication in cell-free massive {MIMO},''
  \emph{arXiv preprint arXiv:2305.12523}, 2023.

\bibitem{9842350}
Y.~Huang, Y.~Fang, X.~Li, and J.~Xu, ``Coordinated power control for network
  integrated sensing and communication,'' \emph{IEEE Trans. on Veh. Technol.},
  vol.~71, no.~12, pp. 13\,361--13\,365, 2022.

\bibitem{9124713}
X.~Liu and {Et. al.}, ``Joint transmit beamforming for multiuser {MIMO}
  communications and mimo radar,'' \emph{IEEE Trans. on Sig. Process.},
  vol.~68, pp. 3929--3944, 2020.

\bibitem{demirhan2023cell}
U.~Demirhan and A.~Alkhateeb, ``Cell-free {ISAC MIMO} systems: Joint sensing
  and communication beamforming,'' \emph{arXiv preprint arXiv:2301.11328},
  2023.

\bibitem{art:su2021}
N.~Su, F.~Liu, and C.~Masouros, ``Secure radar-communication systems with
  malicious targets: Integrating radar, communications and jamming
  functionalities,'' \emph{IEEE Trans. on Wireless Commun.}, vol.~20, no.~1,
  pp. 83--95, 2021.

\bibitem{10104579}
------, ``Sensing-assisted physical layer security,'' in \emph{WSA `I\&' SCC
  2023; 26th International ITG Workshop on Smart Antennas and 13th Conference
  on Systems, Communications, and Coding}, 2023, pp. 1--6.

\bibitem{art:ren}
Z.~Ren, L.~Qiu, and J.~Xu, ``Optimal transmit beamforming for secrecy
  integrated sensing and communication,'' in \emph{ICC 2022-IEEE International
  Conference on Communications}.\hskip 1em plus 0.5em minus 0.4em\relax IEEE,
  2022, pp. 5555--5560.

\bibitem{dasilva2023privacy}
I.~W.~G. da~Silva, D.~P. Osorio, and M.~Juntti, ``Privacy performance of {MIMO}
  dual-functional radar-communications with internal adversary,'' \emph{arXiv
  preprint arXiv:2302.06253}, 2023.

\bibitem{art:paperRFVLC}
I.~W.~G. da~Silva, E.~E.~B. Olivo, M.~Katz, and D.~P.~M. Osorio, ``Secure
  precoding and user association for multiuser hybrid {RF/VLC} systems,''
  \emph{TechRxiv preprint:10.36227/techrxiv.20412177.v2}, 2023.

\bibitem{7405344}
Q.~Shi, M.~Razaviyayn, M.~Hong, and Z.-Q. Luo, ``{SINR} constrained beamforming
  for a {MIMO} multi-user downlink system: Algorithms and convergence
  analysis,'' \emph{IEEE Trans. on Signal Proc.}, vol.~64, no.~11, pp.
  2920--2933, 2016.

\bibitem{art:em}
M.~Haugh, ``The {EM} algorithm,''
  \url{http://www.columbia.edu/~mh2078/MachineLearningORFE/EM_Algorithm.pdf},
  2015.

\bibitem{9882359}
Y.~Zhang, J.~Sun, J.~Xue, G.~Y. Li, and Z.~Xu, ``Deep expectation-maximization
  for joint {MIMO} channel estimation and signal detection,'' \emph{IEEE Trans.
  on Sig. Process.}, vol.~70, pp. 4483--4497, 2022.

\end{thebibliography}
\end{document}